\documentclass[vecphys]{svmult}

\usepackage{makeidx}         
\usepackage{graphicx}        
\usepackage{multicol}        
\usepackage[bottom]{footmisc}

\makeindex

\begin{document}

\title*{Aspects of non leptonic $B_s$ decays}

\author{R. Ferrandes}

\institute{Phys. Dept. University of Bari and INFN, via Orabona 4,
I-70126 Bari, Italy \texttt{rossella.ferrandes@ba.infn.it}}

\maketitle

Non perturbative strong interaction effects make difficult a
theoretical description of non leptonic weak decays of hadrons; so
it is relevant the possibility to obtain predictive
parametrizations of decay amplitudes. It is noticeable that the
widths  of a set of two-body $B_s$ transitions can be predicted
using the symmetries of QCD and available information on  $B$
decays \cite{paper}. We consider in particular a class of decay
modes induced by the transitions $b\rightarrow c\overline{u}q$
($q=d$ or $s$) which are collected in Table \ref{tab1}. These
quark transitions are described by the following effective
hamiltonian, obtained by a renormalization group evolution from
the electroweak scale down to $\mu=m_b$:
$H=\frac{G_F}{\sqrt{2}}V_{cb}V^*_{ud}[c_1(\mu)Q_1+c_2(\mu)Q_2]$,
 where $Q_1=(\overline{c}_i b_i)_{V-A}(\overline{d}_j
u_j)_{V-A}$ and $Q_2=(\overline{c}_i b_j)_{V-A}(\overline{d}_j
u_i)_{V-A}$.

The most popular approach to compute hadronic matrix element of
four quark operators is based on naive factorization \cite{fact},
consisting in factorizing them in current matrix elements
determined in terms of meson decay constants and semileptonic form
factors. This approach has some troubles: firstly, the Wilson
coefficient $c_i(\mu)$ depend on renormalization scale $\mu$,
while form factors, decay constants and physical amplitudes are
scale independent. Moreover, the amplitudes are real, so that
strong phases are neglected; finally, annihilation amplitudes are
predicted to be tiny. Instead, data point to sizeable strong
phases and to non negligible annihilation contributions, as we
shall see.

An alternative approach is a model independent analysis based on
flavor symmetry and experimental data. The key observation is that
the various $B_s$ decay modes are governed, in the $SU(3)_F$
limit, by few independent amplitudes \cite{Zeppenfeld:1980ex} that
can be constrained, both in moduli and in phase differences, from
corresponding $B$ decay processes.
 Since $\overline{B}\rightarrow DP$ decays induced by $b \to c \bar u d(s)$ transitions involve a weak Hamiltonian
transforming as a flavor octet, using the notation $T^{(\mu)}_\nu$
for the $\nu=(Y,I,I_3)$ component of an irreducible tensor
operator of rank $(\mu)$ \cite{deSwart}, one can write:
$H_W=V_{cb}V^*_{ud}T^{(8)}_{0-1-1}+V_{cb}V^*_{us}T^{(8)}_{-1-\frac{1}{2}-\frac{1}{2}}$.
When combined with the initial $\overline{B}$ mesons, which form a
$(3^*)$-representation of SU(3), this leads to $(3^*)$, $(6)$ and
$(15^*)$ representations, which are also those formed by the
combination of the final octet light pseudoscalar meson and
triplet D meson. Therefore, using the Wigner-Eckart theorem, the
decay amplitudes can be written as linear combinations of three
reduced amplitudes $\langle \phi^{(\mu)}|O^{(8)}|B
^{(3^*)}\rangle$, with $\mu=3^*,6,15^*$. By appropriate linear
combinations of these amplitudes one can obtain a correspondence
with the color suppressed, color enhanced and W-exchange diagrams,
$C$, $T$ and $E$, respectively, as in Table \ref{tab1}. The
transition in the $SU(3)$ singlet $\eta_0$ involves another
amplitude $D$ in principle not related to the previous ones. The
$SU(3)$ representation  for  $B$ decays is reported in Table
\ref{tab1}.
\begin{table}
\caption{$SU(3)$ amplitudes for $B^-$, $\overline{B}^0$ and
$\overline{B}_s^0$ decays to $D_{(s)} P$ ($P$ is a light
pseudoscalar meson), induced by $b \to c \overline{u}d(s)$
transitions. The predicted $\overline{B}_s^0$ branching fractions
are also reported.}
 \label{tab1}
 \begin{center}

 \begin{tabular}{lcc||lcc}
 \hline\noalign{\smallskip}
 $B^-, \overline{B}^0 $&  amplitude & BR$_{exp}$ $(10^{-4})$ & $\overline{B}_s^0$ &  amplitude & BR$_{th}$ $(10^{-4})$ \\
 \noalign{\smallskip}\hline\noalign{\smallskip}
 $ D^0\pi^-$ & $V^*_{ud}V_{cb}\,\, (C+T)$&  $49.8\pm2.9$  & $ D_s^+\pi^-$ & $V^*_{ud}V_{cb}\,\, T$&  $29\pm6$ \\
 $ D^0\pi^0$ & $V^*_{ud}V_{cb}\frac{(C-E)}{\sqrt{2}}$& $2.91\pm0.28$  & $ D^0\overline{K}^0$ & $V^*_{ud}V_{cb}\,\,C$&  $8.1\pm1.8$ \\
 $ D^+\pi^-$ & $V^*_{ud}V_{cb}\,\, (T+E)$& $27.6\pm2.5$  & $ D^0 \eta_8$ &$V^*_{us}V_{cb}\frac{(2 C-E)}{\sqrt{6}}$ &  \\
 $ D^+_sK^-$ & $V^*_{ud}V_{cb}\,\,  E$& $0.38\pm0.13$   & $ D^0 \eta_0$ &$V^*_{us }V_{cb}\,\, D$ &  \\
 $ D^0 \eta_8$ & $ -V^*_{ud}V_{cb}\frac { (C+ E)}{\sqrt{6}}$&  & $ D^0 \eta$ & &$0.21\pm0.12 $  \\
 $ D^0  \eta_0$ & $V^*_{ud}V_{cb}\,\,  D$&    & $D^0 \eta^\prime$ & & $0.10\pm0.08 $\\
 $ D^0 \eta$ & & $2.2\pm0.5$  & $D^0\pi^0$ & $- V^*_{us}V_{cb}\frac{ E}{\sqrt{2}}$&  $0.010\pm0.003$ \\
 $ D^0  \eta^\prime$ & &   $1.7\pm0.4$   & $ D^+\pi^-$ &$V^*_{us}V_{cb}\,\, E$ & $0.020\pm0.006$ \\
 $ D^0K^-$ & $V^*_{us}V_{cb}\,\, (C+T)$ &$3.7\pm0.6$  & $ D_s^+ K^-$ &$V^*_{us}V_{cb}\,\, (T+E)$ & $1.8\pm0.3$ \\
 $ D^0\overline{K}^0$ & $V^*_{us}V_{cb}\,\, C$  & $0.50\pm0.14$ & & &\\
   $ D^+K^-$ & $V^*_{us}V_{cb} \,\, T$ &$2.0\pm0.6$ & & & \\
 \noalign{\smallskip}\hline
 \end{tabular}
 \end{center}
\end{table}

\begin{figure}[ht]
\begin{center}
\includegraphics[width=0.40\textwidth] {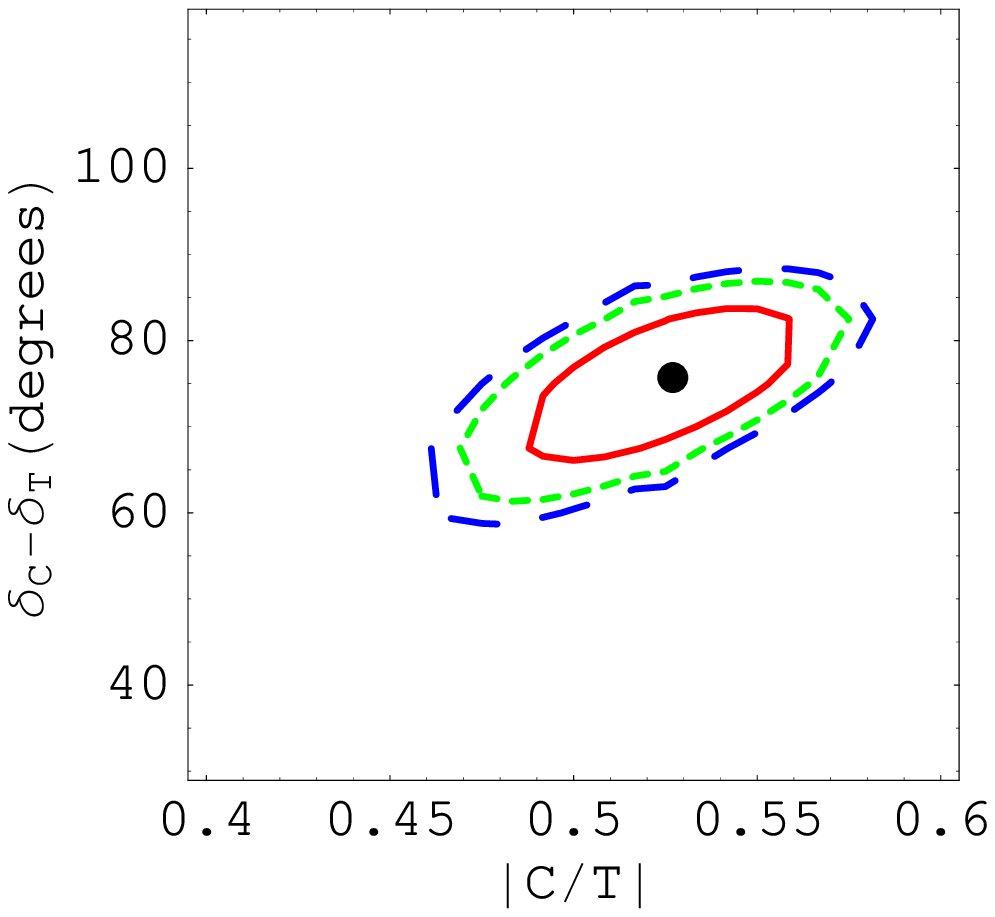} \hspace*{0.3mm}
\includegraphics[width=0.40\textwidth] {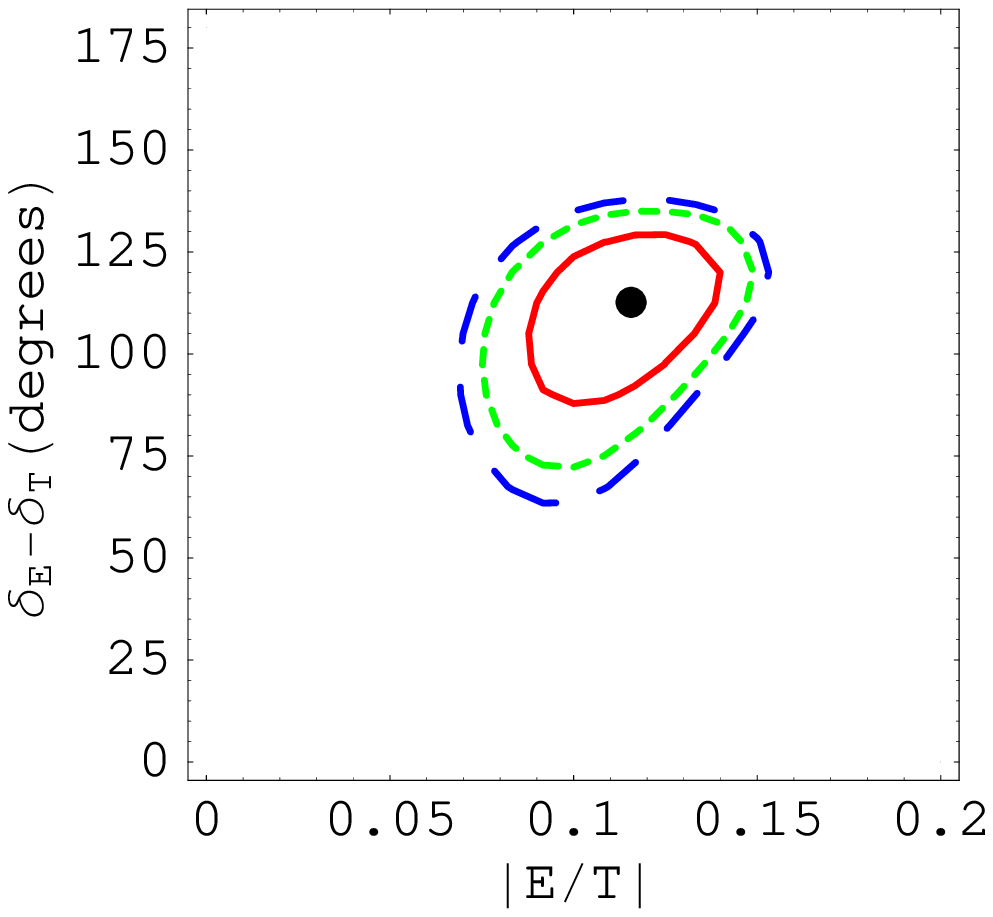}
\vspace*{0mm} \caption{\baselineskip=15pt Ratios of SU(3)
amplitudes   obtained from $B$ data in Table \ref{tab1}.  The
contours correspond to the confidence level of $68\%$ (continuous
line), $90\%$ (dashed line) and $95\%$ (long-dashed line); the
dots  show the result of the fit.} \label{fig:BDM}
\end{center}
\end{figure}

We note that $\bar B \to D_s K$ only
 fixes the modulus of $E$, which is not small at odds with the expectations by factorization, where
$W$-exchange processes are suppressed by ratios of decay constants
and form factors and are usually considered to be negligible. What
can be done is to use all the information on $\bar B \to D \pi,
D_s K$ and $D K$ (7 experimental data) to determine $T$, $C$ and
$E$ (5 parameters).  A similar strategy has been recently adopted
in \cite{kim}. Noticeably, the combined experimental information
is enough accurate to tightly determine the ranges of variation
for all these quantities. In  fig. \ref{fig:BDM} we have depicted
the allowed regions in the $C/T$ and $E/T$ planes, obtained fixing
the other variables to their fitted values, with the corresponding
confidence levels. It is worth noticing that
 the phase differences between the various amplitudes
are close to be maximal;  this signals  sizeable deviation from
naive (or  generalized) factorization, provides contraints to
QCD-based approaches proposed to evaluate non leptonic $B$ decay
amplitudes \cite{Beneke:2000ry, Bauer:2001cu,Keum:2003js}
 and points towards large long-distance effects in $C$ and $E$
 \cite{violations}. We obtain $|\frac{C}{T}|=0.53 \pm 0.10$,  $|\frac{E}{T}|=0.115\pm0.020$,
$\delta_C-\delta_T=(76\pm12)^\circ$   and
$\delta_E-\delta_T=(112\pm46)^\circ$.

With the results for the amplitudes  we can determine a number of
$B_s$ decay rates, and   the predictions are collected in Table
\ref{tab1}. The uncertainties in the predicted rates are small; in
particular, the $W$-exchange induced processes   $\overline{B}_s^0
\to D^+ \pi^-, D^0 \pi^0$ are precisely estimated  \cite{aleksan}.
The predicted ratio $\frac{\Gamma(B_s \to D_s^- \pi^+)}{\Gamma(B^0
\to D^- \pi^+)}=1.05\pm 0.24$ can be compared to the experimental
value $1.32\pm 0.18 \pm 0.38$ recently obtained by CDF
Collaboration \cite{CDF}.

We have performed an analogous analysis for $\bar B \to D_{(s)}
V,$ $D_{(s)}^* P$ decays, for which the same $SU(3)$ decomposition
holds, with amplitudes $T'$, $C'$, $E'$. $B$ decay data are
collected in Table \ref{tab3}, including the recently observed
W-exchange mode $\overline{B}^0 \to D_s^* K^-$ \cite{babar},
together with the predictions for $B_s$ decays.

$SU(3)_F$ breaking terms can modify our predictions. Those effects
in general cannot be reduced to well defined and predictable
patterns without new assumptions. Their parametrization  would
introduce additional  quantities \cite{Gronau:1995hm}  that  at
present cannot be sensibly bounded since  their effects seem to be
smaller than  the experimental uncertainties. It will be
interesting to investigate their role when the $B_s$ decay rates
will be measured and more precise $B$ branching fractions  will be
available.

\index{paragraph}

\begin{table}
\caption{Experimental branching fractions of $\bar B \to D_{(s)}
V,$ $D_{(s)}^* P$ decays and predictions for $\overline{B}_s^0 $
decays.}
    \label{tab3}
    \begin{center}
    \begin{tabular}{lc||lc}
  \hline \noalign{\smallskip}
$B^-, \overline{B}^0 $ &  BR$_{exp}$ $(10^{-4})$ & $\overline{B}_s^0$ & BR$_{th}$ $(10^{-4})$\\
\noalign{\smallskip} \hline \noalign{\smallskip}
$ D^0\rho^-$ &  $134\pm18$&$ D _s^+ \rho^{-}$ &   $72\pm 35$ \\
$ D^0\rho^0$ &  $2.9\pm1.1$&$ D^0\overline{K}^{*0}$ & $9.6\pm2.4$ \\
$ D^+\rho^-$    &  $77\pm13$& &  \\
$ D_s K^{*-}$    &  $<  \,9.9$& & \\

 $ D^0 K^{*-}$ &  $6.1\pm2.3$&$ D^0 \rho^{0}$ &   $0.28\pm 1.4$ \\
$ D^0 \bar K^{*0}$ &  $0.48\pm0.12$&$ D^+ \rho^-$ & $0.6\pm 2.8$ \\
$ D^+ K^{*-}$    & $3.7\pm1.8$&
$ D_s^+K^{*-}$ &  $4.5\pm3.1$ \\
\hline
$D^{*0}\pi^-$ &  $46\pm4$&$ D _s^{*+} \pi^{-}$ &   $32\pm 2$ \\
$ D^{*0}\pi^0$ &  $2.7\pm0.5$&$D^{*0}\overline{K}^{0}$ & $4.7\pm2.2$ \\
$D^{*+}\pi^-$    &  $27.6\pm2.1$& &  \\
$ D_s^{*} K^{-}$    &  $0.20\pm0.06$& & \\

 $ D^{*0} K^{-}$ &  $3.6\pm1.0$&$ D^{*0} \pi^{0}$ &   $0.0057\pm 0.0017$ \\
 &  &$ D^{*+} \pi^-$ & $0.0115\pm 0.0035$ \\
$ D^{*+} K^{-}$    & $2.0\pm0.5$&
$ D_s^{*+}K^{-}$ &  $1.3\pm0.2$ \\
\noalign{\smallskip} \hline
      \end{tabular}
   \end{center}
  \end{table}

  \vspace*{1cm}
\noindent {\bf Acnowledgments}\\
\noindent I thank P. Colangelo for collaboration and F. De Fazio
for useful discussions.

%
%

%
%



\printindex
\end{document}